\documentclass[a4paper,11pt]{article}
\usepackage{pos}

\title{Angular Distributions in Rare $\mathbf{b}$ Decays}
\ShortTitle{Angular Distributions in Rare ${b}$ Decays}

\author*[a]{Thorsten Feldmann}

\affiliation[a]{Theoretische Physik 1, Universit\"at Siegen, Walter-Flex-Stra\ss{}e 3, D-57068 Siegen, Germany}

\note[]{Preprint P3H-21-003, SI-HEP-2021-01}

\emailAdd{thorsten.feldmann@uni-siegen.de}

\abstract{We give a brief overview of phenomenological developments in the analysis of angular observables for exclusive decay modes of $B$ mesons and $\Lambda_b$ baryons, with focus on recent results and some important aspects related to the theoretical background (which mostly concern the 
	treatment of hadronic uncertainties).}

\FullConference{%
  BEAUTY2020\\
  21-24 September 2020\\
  Kashiwa, Japan (online)}

\begin{document}
\maketitle

\section{Introduction}

In recent years, angular observables in rare $b$-quark decays have gained a lot of attention in theoretical and phenomenological studies that aim to find indirect hints on physics beyond the Standard Model (SM). A review on the benefits of 
angular observables for global fits of Wilson coefficients in the weak effective Hamiltonian, and the  role of the associated 
''anomalies''  observed in experimental measurements -- including hints for lepton-flavour-universality violation -- has been presented by Sébastien Descotes-Genon at last year's conference
of this series \cite{Descotes-Genon:2020ijt}. More references as well as summaries of the experimental prospects at LHCb and Belle~II -- including the ''golden decay channels'' $B \to K \mu^+ \mu^-$ and $B\to K^* \mu^+\mu^-$ --  can be found, for instance, in \cite{Bediaga:2012py,Kou:2018nap}.
The common goal of these investigations is to further improve the experimental constraints on the relevant Wilson coefficients $C_{9,10,9',10',\ldots}$ that describe the short-distance physics in $|\Delta B|=|\Delta S|=1$ transitions.
Here, the angular observables in the individual exclusive decay modes give detailed information on the decay dynamics, which can be optimized such that systematic experimental and theoretical uncertainties cancel to some extent. To make full use of the existing theoretical and experimental results, a careful statistical analysis
is needed. One the one hand, this concerns the proper discrimination between SM and NP interpretation of the data. 
One the other hand, one needs to take into account parametric and systematic uncertainties, including their correlations. Details about the procedure and the status of such global fits can be found in the contribution by Javier Virto to this conference \cite{Virto:2020}.

\section{Theoretical toolbox}

The Wilson coefficients $C_i(\mu)$ in the weak effective Hamiltonian can be precisely calculated in the SM or in specific NP models, and the scale-dependence is 
controlled by the renormalization-group equations in the low-energy effective theory. On the other hand, the evaluation of hadronic matrix elements of the effective operators requires additional approximations. Intuitively, these matrix elements naively factorize into hadronic form factors and purely leptonic amplitudes, where the former can be estimated to quite good accuracy from light-cone sum rules at large momentum transfer or lattice-QCD analyses at small momentum transfer. Corrections to this picture in radiative and rare semileptonic decays arise from contributions of hadronic operators, where the real or virtual photon is emitted within the long-distance hadronic transition. 
Some of these corrections can be absorbed into ''effective'' Wilson coefficients $C_9^{\rm eff}(q^2)$ and $C_7^{\rm eff}$, but additional non-factorizable effects remain where hadronic form factors are not sufficient to provide the relevant non-perturbative information. At low recoil ($q^2\gtrsim 16$~GeV$^2$) such effects can be systematically studied in the framework of heavy-quark effective theory (HQET). At large recoil ($q^2 \lesssim 6$~GeV$^2$), non-factorizable effects can be addressed in the framework of QCD factorization (QCDF) or soft-collinear effective theory (SCET), respectively. In both cases, this involves a simultaneous expansion in the strong coupling and inverse powers of the heavy-quark mass which leads to the appearance of new hadronic input functions. To combine theoretical results from different regions of phase space and experimental information on hadronic amplitudes -- including possible resonances in the $q^2$-spectrum -- one can exploit dispersion relations which are associated with analyticity and unitarity properties of the hadronic correlators under concern.

\section{Decays of $B$ mesons}

The benefit of optimized angular observables for NP searches 
in $B \to K^* \ell^+\ell^-$ decays has been emphasized in a 
number of theoretical and phenomenological analyses, for an imcomplete list, see e.g.\ \cite{Matias:2012xw,Bobeth:2012vn,Descotes-Genon:2013vna}.
Fitting the angular decay distributions in combination 
with the signals for violation of lepton-flavour 
universality in $b\to s\ell^+\ell^-$ transitions,
 indicate that the deviation of 
the Wilson coefficient $C_9$ from its SM value can be as 
large as 25\%. This observation has triggered  
phenomenological studies for many other decay modes; recent examples are a time-dependent angular analysis  for
 $ B_d^- \to K_S \ell^+\ell^-$ \cite{Descotes-Genon:2020tnz}
or an angular analysis for $B_s \to f_2 (\to KK ) \mu^+ \mu^-$
\cite{Rajeev:2020aut}. 
Experimental prospects have been discussed by Adl\`ene Hicheur at this conference \cite{Hicheur:2020}.

\paragraph{Decay amplitudes for $\mathbf{B\to K^{*}\ell^+\ell^-}$: }

\begin{figure}
	\centering
	\includegraphics[height=0.32\linewidth]{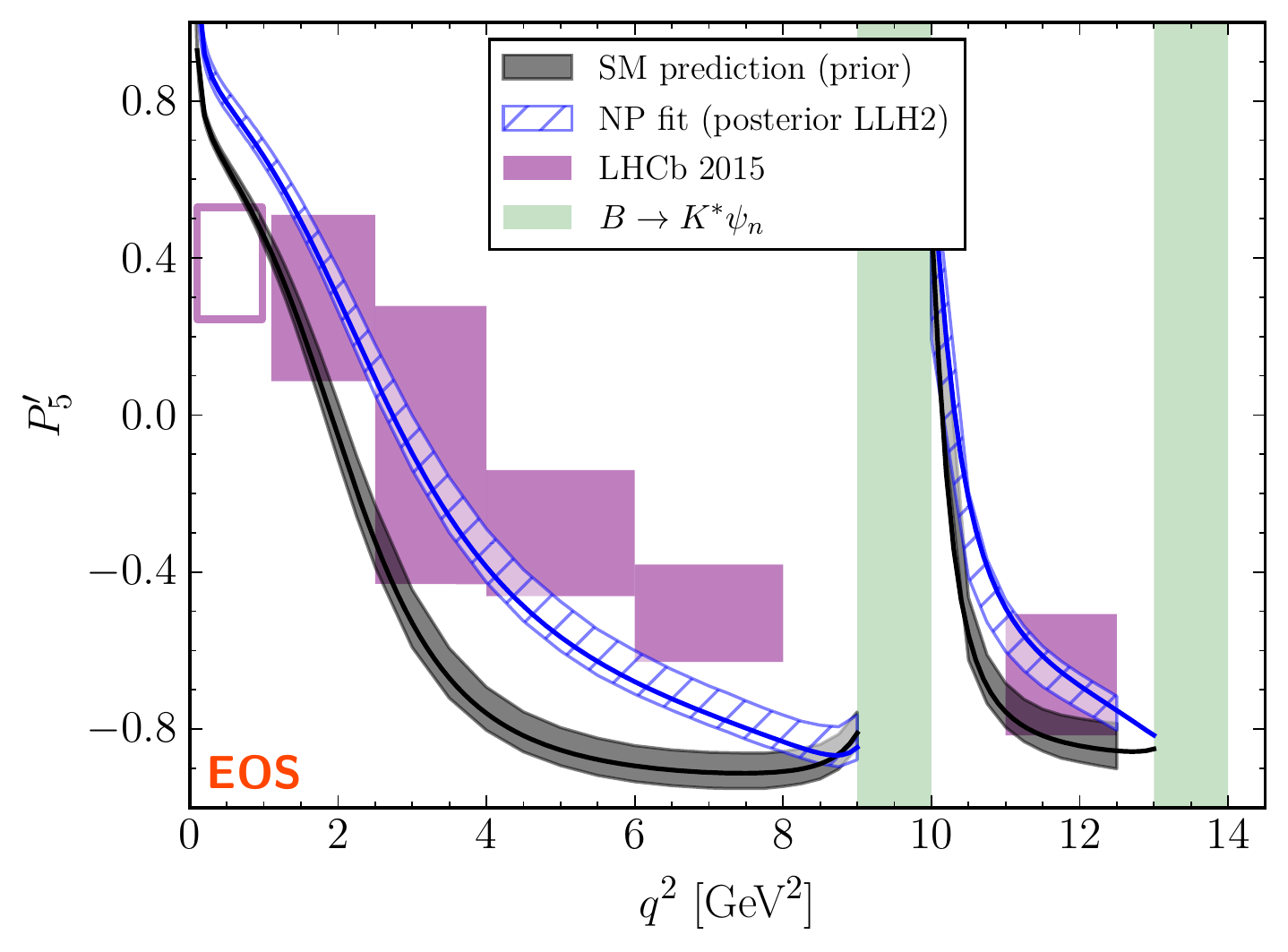}
	\qquad 
	\includegraphics[height=0.31\linewidth]{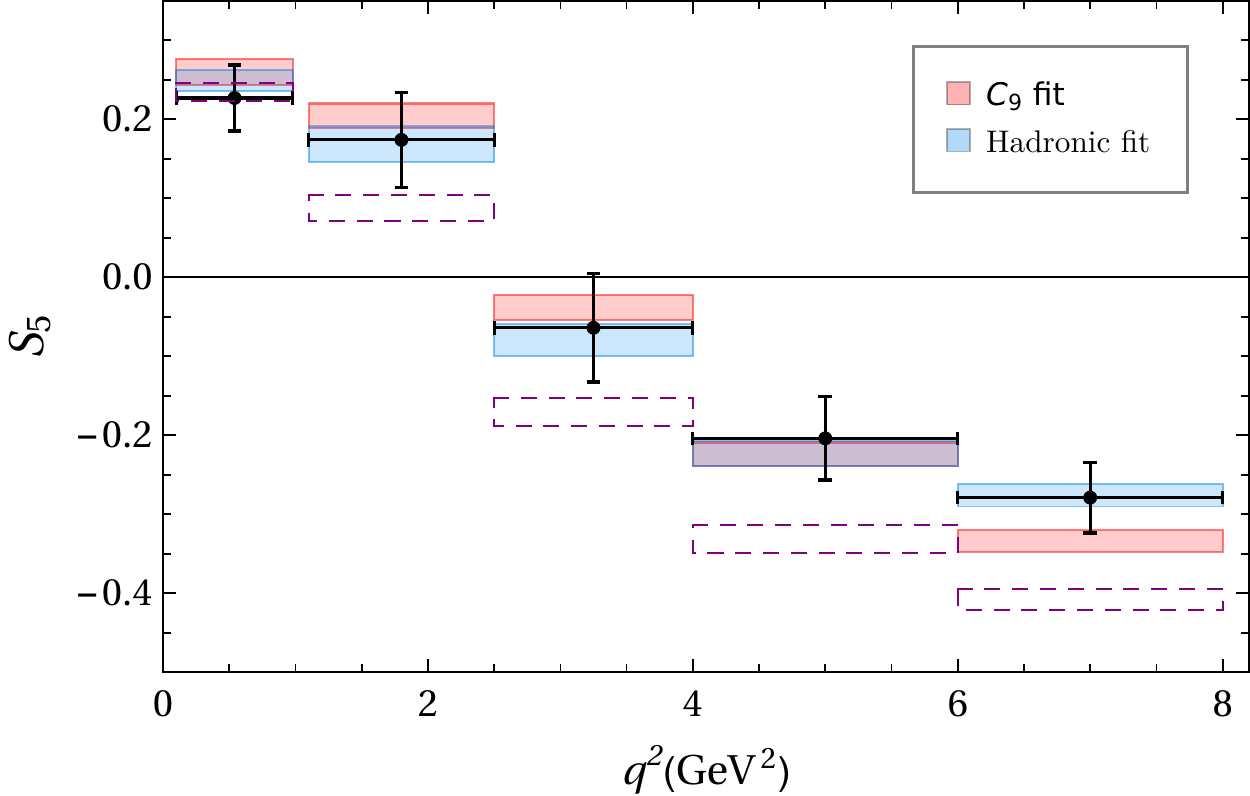}
	\caption{Phenomenological analyses of the angular observable $P_5'$ (respectively $S_5$). Left: From
	 \cite{Bobeth:2017vxj} based on (\ref{zexp}), where the NP fit allows for deviations in the Wilson coefficient $C_9$. Right: From \cite{Hurth:2020rzx} where the hadronic fit correspond to an agnostic treatment of theory uncertainties.} 
	\label{fig:s5lowq2fits}
\end{figure}

As an example, let us briefly discuss the state of the art concerning the theoretical description of $B \to K^*\ell^+\ell^-$ decay amplitudes (which feed into the predictions for angular observables).
Following the notation in \cite{Bobeth:2017vxj}, the general decomposition of the $B \to K^*\ell^+\ell^-$ transversity amplitudes in the SM can be written as
\begin{equation}
	A_\lambda^{L,R} \ \propto \ 
	{(C_9\pm C_{10})} \, {{\cal F}_\lambda(q^2)} 
	+\frac{2 M_B^2}{q^2} \left[ 
	\frac{m_b \, {C_7}}{M_B} \, {{\cal F}_\lambda^T(q^2)} 
	- 16\pi^2  \, {{\cal H}_\lambda(q^2)} \right]
\end{equation}
with $\lambda=\perp,\parallel,0$. Here factorizable 
hadronic effects are encoded in form-factor functions 
${\cal F}_\lambda^{(T)}(q^2)$, while non-factorizable hadronic
effects enter through the helicity- and $q^2$-dependent functions ${\cal H}_\lambda(q^2)$ which contain the leading-order quark-loop diagrams and perturbative and non-perturbative corrections. Theoretical calculations in QCDF or SCET constrain the functions ${\cal H}_\lambda$ for values of momentum transfer far below the partonic $c\bar c$-threshold, $q^2 \ll 4m_c^2$, extending to the (unphysical) space-like region ($q^2 < 0$). On the other hand,
the resonant contributions to ${\cal H}_\lambda$ from 
the low-lying charmonium states via $B \to J/\psi K^*$ and $B\to \psi(2S) K^*$ can be  determined experimentally.
The central idea is to use the conformal mapping 
\begin{equation}
			q^2 \mapsto  z(q^2)\equiv \frac{\sqrt{t_+-q^2}-\sqrt{t_+-t_0}}{\sqrt{t_+-q^2}+\sqrt{t_+-t_0}}
		\end{equation}
with $t_+=4M_D^2$ denoting the open-charm threshold,
and an auxiliary parameter $t_0$ which can be optimized
to  minimize the absolute value $|z|$ in the region of interest. 
Then, the non-factorizable effects can be expanded in $z$,
\begin{eqnarray}
	&&	{\cal H}_\lambda(z) = 
		\frac{1-z \, z^*_{J/\psi}}{1-z_{J/\psi}}
	\, 
	\frac{1-z \, z^*_{\psi(2S)}}{z-z_{\psi(2S)}}
	\, {{\cal F}_\lambda(z)} \, \sum_{k=0}^K\, { 
		\alpha_k^{(\lambda)}}
	\, z^k 
	\label{zexp}
\end{eqnarray}
Here one concentrates on the charmful operators $O_{1,2}^{(c)}$, such that the first two prefactors correctly reproduce the analytic structure from the considered charmonium resonances. 
The normalization to the form-factor functions ${\cal F}_\lambda(z)$ has been introduced for convenience. One is thus left with the parameters $a_k^{(\lambda)}$ which have to be fitted to experimental data and theoretical predictions for a given truncation $K$. In this way one can set up a well-defined framework for a proper analysis of hadronic uncertainties and 
a reliable extraction of SM Wilson coefficients from the angular observables in $B \to K^*\mu^+\mu^-$, including experimental results below and inbetween the $J/\psi$ and $\psi'$ resonances.
As an example, we present a comparison of the posterior distributions for the angular observable $P_5'$ in the SM and in a NP scenario with extra contributions to the Wilson coefficient $C_9$ on the left-hand side of Figure~\ref{fig:s5lowq2fits}.
 More details can be found in \cite{Bobeth:2017vxj}.

The question of how to disentangle NP effects from non-factorizable hadronic uncertainties in the angular observables from $B \to K^*\mu^+\mu^-$ decays has also been re-addressed recently in \cite{Hurth:2020rzx} (for earlier work, see also 
\cite{Ciuchini:2015qxb}). 
In that work the hadronic uncertainties are treated in an agnostic manner. It is found that 
the collective effects of small deviations in individual hadronic parameters can also lead to a reasonable description of the experimental data on angular observables like $S_5$, see the right-hand side of Figure~\ref{fig:s5lowq2fits}.

\section{Decays of $\Lambda_b$ baryons}

Rare semileptonic decays of $\Lambda_b$ baryons provide 
a large number of angular observables which often result in 
complementary sensitivity to different Dirac structures 
entering the effective Hamiltonian in generic NP extensions.
In particular, depending on the experimental setup, 
the $\Lambda_b$ can be produced with non-vanishing polarization
(which can be tested in the angular distributions themselves).
One expects to find  similar deviations from the SM in the baryonic decay modes as in $B \to K^*\mu^+\mu^-$.

From the theoretical point of view one has to keep in mind 
that the light spectator system in the $\Lambda_b$ has the 
quantum numbers of a diquark. This results in independent and different types of hadronic uncertainties compared to the $B$-meson case. While $\Lambda_b$ transition form factors are available from lattice-QCD studies, the understanding 
of non-factorizable spectator-dependent effects is still rather poor.

\paragraph{The decay $\mathbf{\Lambda_b \to \Lambda(\to p\pi)\ell^+\ell^-}$:}

The $\Lambda_b \to \Lambda$ transition is described by 
10 independent form factors which can be conveniently defined in
a helicity basis (see e.g.\ \cite{Feldmann:2011xf}). In the low-recoil limit (HQET) the number of independent form factors reduces to 2; at large recoil (SCET) this further reduces to one single form factor. Quantitative predictions from lattice QCD 
can be found in \cite{Detmold:2016pkz} for low and moderate recoil energies. The results are extrapolated to large recoil using a $z$-expansion in an analogous way as described above.
A detailed theoretical discussion of the angular analysis for $\Lambda_b \to \Lambda(\to p\pi)\ell^+\ell^-$ can be found in
\cite{Boer:2014kda} (see also \cite{Gutsche:2013pp}). 
In the unpolarized case, one encounters 10 angular observables
which -- besides of the Wilson coefficients and (generalized) form factors -- now also depend on 
a parity-violating decay parameter $\alpha$ for weak $ \Lambda\to N \pi$ decays.
As a result one encounters additional forward-backward asymmetries 
(as compared to the $B \to K^*$ mode) which are sensitive to 
other and independent combinations of Wilson coefficients.
By constructing optimized angular observables one can 
indeed extract complementary information on $b \to s \ell^+\ell^-$ transitions from $\Lambda_b \to \Lambda$ decays.
The kinematics of polarized $\Lambda_b$ decays is described by 
five angles which allow to identify 24 \emph{additional} angular observables \cite{Blake:2017une}.

\begin{figure}
	\centering
	\includegraphics[height=0.44\linewidth]{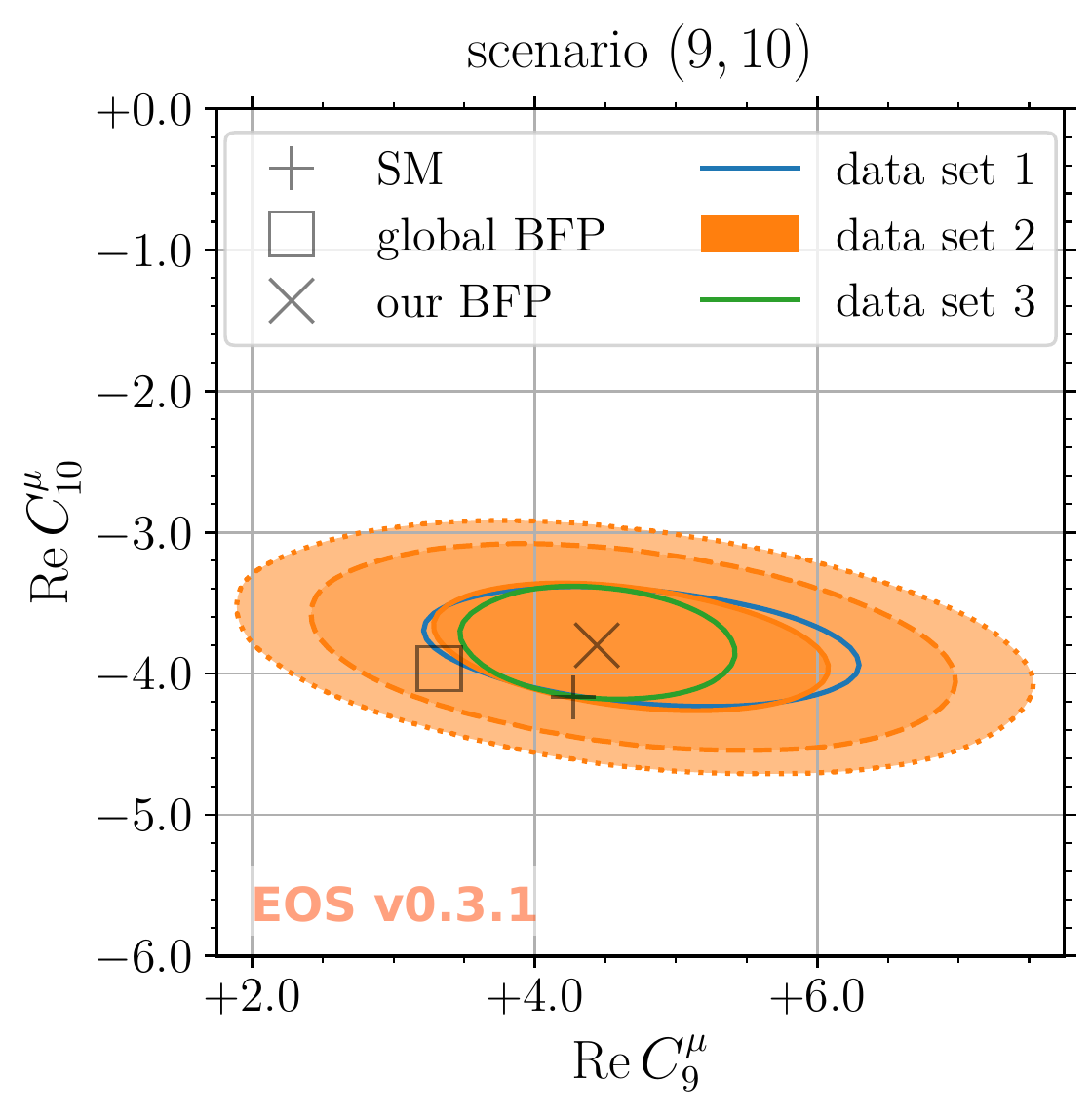}
\quad 
	\includegraphics[height=0.39\linewidth]{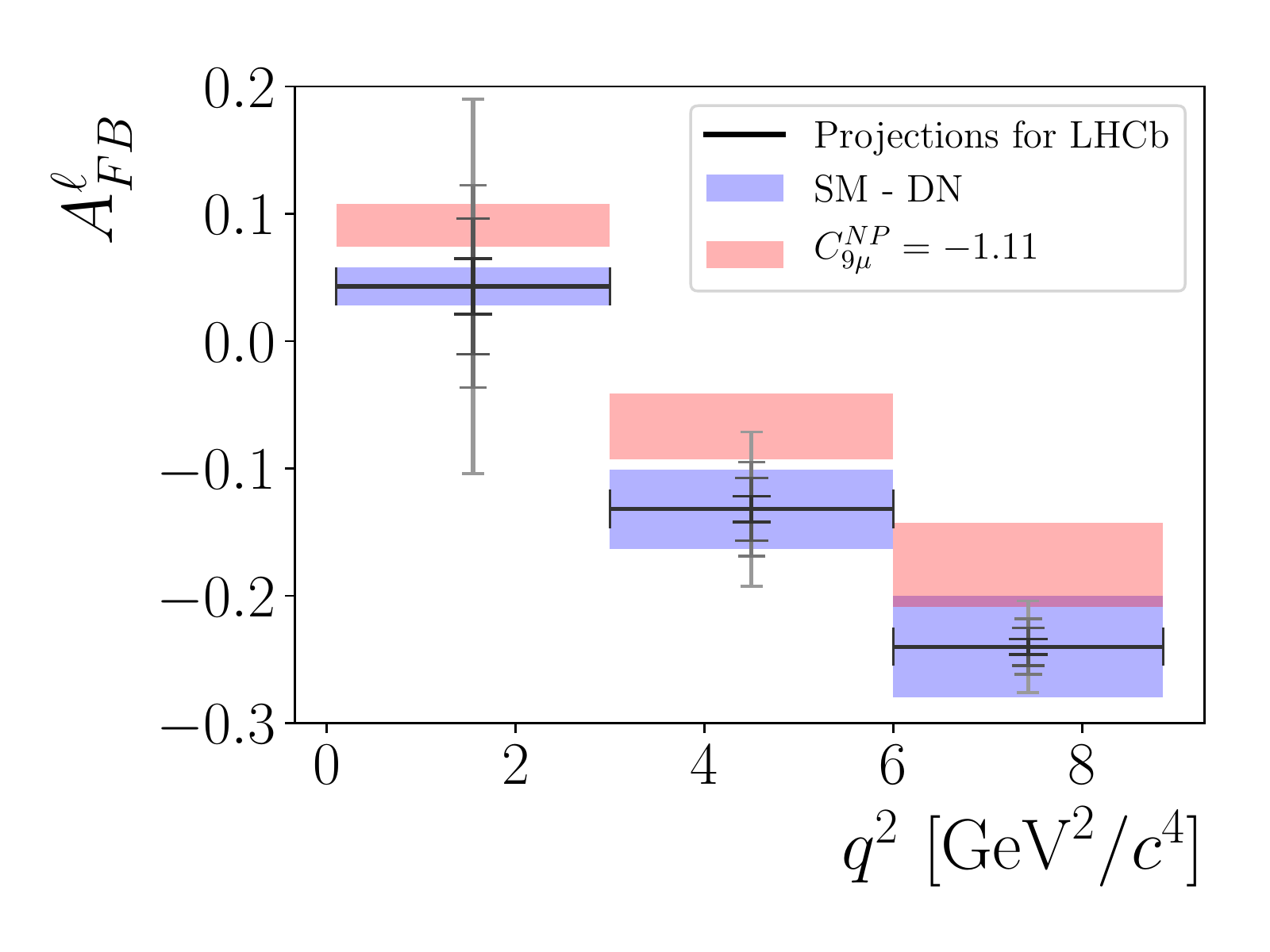}
	\caption{Left: Fit result of angular analysis for $\Lambda_b \to \Lambda(\to N\pi) \ell^+\ell^-$ from \cite{Blake:2019guk}. Right: Projected sensitivity to see NP in the lepton forward-backward asymmetry in $\Lambda_b \to \Lambda^*(\to K\pi) \ell^+\ell^+$ from \cite{Amhis:2020phx}.}
	\label{fig:sensitivityafb10000twithaccalltheorysmallbins}
\end{figure}

An updated phenomenological analysis of 
$\Lambda_b \to \Lambda(\to p\pi^-)\mu^+\mu^-$
has recently been performed in \cite{Blake:2019guk}.
Notably, this includes new results for the parity-violating parameter $\alpha$, the measurement of the complete
set of angular observables from LHCb \cite{Aaij:2018gwm}, constraints from the time-integrated branching fraction for 
$B_s\to\mu_+\mu_-$, as well as
an updated value for the $\Lambda_b$ fragmentation function 
and the resulting branching fraction for $\Lambda_b \to J/\psi \Lambda$ which is used as a normalization in the LHCb measurement of $\Lambda_b \to \Lambda\mu^+\mu^-$. Among the main results of this analysis are: the $\Lambda_b$ polarization at LHCb is compatible with zero, $|P_{\Lambda_b}| \leq 11\% (@95\%)$;
the angular distribution is compatible with the SM;
one finds a similarly good fit with NP in $C_9$ only, leading 
to $C_9 = 4.8 \pm 0.8$, and a slightly better fit for NP in $C_{9,10}$ with $C_9 = 4.4 \pm 0.8$,  $C_{10}=-3.8\pm 0.3$.
For the latter case the fit result is illustrated on the left 
of Figure~\ref{fig:sensitivityafb10000twithaccalltheorysmallbins}.
As can be seen, the fit with the baryonic decay is compatible with the NP global fit for $B$-meson decays ($\square$), but also with the SM expectations (+). More details and references to earlier work in the same direction can be found in \cite{Blake:2019guk}.

Another recent analysis of $\Lambda_b \to \Lambda(\to N\pi)\ell^+\ell^-$ 
can be found in \cite{Yan:2019tgn}. In that work the 
full set of semileptonic operators, including Dirac tensor currents, have been taken into account, and also the lepton mass is kept finite to allow applications with $\tau$-leptons in the final state.
As a phenomenological example, the authors compare the SM prediction with a scalar-leptoquark model. However, the updated 
LHCb data \cite{Aaij:2018gwm} were not included in the fit.

\paragraph{Lepton-flavour-violating modes $\mathbf{\Lambda_b \to \Lambda \, \ell_1^+\ell_2^-}$:}

NP models that explain the violation of lepton-flavour universality in $B$-meson decays often also lead to sizable charged lepton-flavour
violation (LFV). In \cite{Das:2019omf} 
LFV decays $b \to s \ell_1^+\ell_2^-$ have 
been analyzed in $\Lambda_b \to\Lambda$ transitions.
A welcome advantage compared to the lepton-flavour conserving modes is the absence of non-factorizable long-distance 
effects.
As LFV is tiny in the SM, any observation of $\Lambda_b \to \Lambda \ell_1^+\ell_2^-$ would be a clear sign of NP.
The analysis in 	\cite{Das:2019omf} 
includes all vector, axial-vector, scalar and pseudo-scalar operators, and evaluates the branching ratio and leptonic FB asymmetry in terms of the angular coefficients.
As a phenomenological benchmark model the authors consider 
a scenario with a vector leptoquark $U_1=(3,1)_{3/2}$, where
the parameter space is constrained by other low-energy observables.
Given the still rather loose constraints on NP from $B_s \to \tau^+ \tau^-$ and $B \to K \tau^+\tau^-$, the authors come to the conclusion that the branching ratios for LFV $\Lambda_b$ decays could potentially be accessible at LHCb.

\paragraph{Decays of $\Lambda_b$ to excited $\Lambda(1520)$:}

In contrast to the above decays, the excited baryon $\Lambda(1520)$ decays by \textit{strong} interaction 
into $pK$ or $nK$. It appears to dominate the $\Lambda_b \to pK^-J/\psi$ phase space around invariant masses $m_{pK}\sim 1.5$~GeV.
As the $\Lambda(1520)$ has spin-parity $J^P=3/2^-$ it can 
provide complementary information on NP in $b \to s \ell^+\ell^-$ 
transitions. On the other hand,
$\Lambda_b \to \Lambda^*$ form factors are more involved on the lattice; very recent results can be found in \cite{Meinel:2020owd}.
Similarly, the theoretical knowledge on the hadronic structure of the $\Lambda(1520)$ entering the non-factorizable contributions to the $\Lambda_b$ decay are very poor. It should also be noted that in the physical phase space the recoil energy is not particularly large and the mass of the excited $\Lambda(1520)$ is not particularly small. Therefore, symmetry relations in HQET or SCET
will receive potentially large corrections (the $\alpha_s$ corrections to the HQET form-factor relations at low recoil have been worked out in \cite{Das:2020cpv}). 

The angular observables for $\Lambda_b \to \Lambda^*(\to N K)\ell^+\ell^-$ have been studied in \cite{Das:2020cpv,Descotes-Genon:2019dbw}.
Compared to  $\Lambda_b \to \Lambda(J/P=1/2^+)$
transitions one has to properly deal with the 
theoretical subtleties concerning the 
quantization of spin-3/2 fields, which, however,
turn out to be irrelevant in the narrow-width approximation. It is thus sufficient to describe the $\Lambda(1520)$ state by 
a conventional Rarita-Schwinger spinor.
The number of independent form factors is increased to 14;		
however, the additional form factors should vanish in the HQET or SCET limit. Similarly, the differential decay rate for unpolarized $\Lambda_b \to \Lambda^* \ell^+\ell^-$ decays is now described in terms of 12 angular coefficients (instead of 10 for $\Lambda_b \to \Lambda$).
Prospects for future NP studies at LHCb based on 
preliminary numerical studies of $\Lambda_b \to \Lambda(1520)(\to N\bar K)\ell^+\ell^-$ from \cite{Descotes-Genon:2019dbw}
have been discussed in \cite{Amhis:2020phx}. It is observed that some of the angular coefficients show reasonable sensitivity to 
NP contributions of the size that is indicated by the current global fits to $B$-meson decays. As an example we show on the right of Figure~\ref{fig:sensitivityafb10000twithaccalltheorysmallbins}
the prospects for future measurements of the leptonic forward-backward asymmetry, comparing the SM expectations and a NP scenario with deviations in the Wilson coefficient $C_9$. (It has to be noted that the estimates in \cite{Descotes-Genon:2019dbw} are still based on simplified models for the form factors.)
In the same analysis it has also been pointed out that the vanishing of the hadronic forward-backward asymmetries (as a consequence of the strong decay of $\Lambda(1520)$) 
could eventually be exploited for the experimental identification 
of $\Lambda(1520)$ candidates in the multi-body final state.

\section{Summary and outlook}

Angular observables in exclusive $b\to s\ell^+\ell^-$ decays contain crucial information on  short- and long-distance dynamics in $b$-hadron decays, and provide the interface between experimental measurements, phenomenological analyses, and theoretical interpretation. There has been a lot of recent progress in reducing hadronic uncertainties (including non-factorizable contributions) from data-driven methods. 
Several model-independent global fits with different 
implementations of SM uncertainties or NP scenarios are available,
which also take into account the interplay with LFU-violating observables.
 As we have seen, new decay modes and observables are proposed for further cross-checks of exclusive $b \to s\ell^+\ell^-$ transitions in the future which, however, also require more sophisticated theory analyses, in particular for baryonic modes. For some of the precision observables one might even have to care about logarithmically enhanced QED corrections which should be addressed properly, 
 see e.g.\ recent work in that direction in \cite{Isidori:2020acz}.


\section*{Acknowledgments}
		
The research of TF is supported by the Deutsche Forschungsgemeinschaft (DFG, German Research Foundation) under grant  396021762 - TRR 257.

\end{document}